# Educational trajectories of graduate students in physics education research

Ben Van Dusen
*School of Education, University of Colorado-Boulder, Boulder, Colorado 80309, USA*

Ramón S. Barthelemy
*Department of Physics, University of Michigan, Ann Arbor, Michigan 48105, USA*

Charles Henderson
*Department of Physics and Mallinson Institute for Science Education, Western Michigan University, Kalamazoo, Michigan 49008, USA*

Physics education research (PER) is a rapidly growing area of PhD specialization. In this article we examine the trajectories that led respondents into a PER graduate program as well as their expected future trajectories. Data were collected in the form of an online survey sent to graduate students in PER. Our findings show a lack of visibility of PER as a field of study, a dominance of work at the undergraduate level, and a mismatch of future desires and expectations. We suggest that greater exposure is needed so PER is known as a field of inquiry for graduates, that more emphasis should be placed on research beyond the undergraduate level, and that there needs to be stronger communication to graduate students about potential careers.

## I. INTRODUCTION

A growing body of literature has begun to explore the lives, working conditions, and academic persistence of members of the Discipline Based Education Research (DBER) community [1–4]. Such research comes on the heels of the National Research Council's report summarizing the state of DBER and suggestions for future growth and development [5]. One suggestion from this report is that the educational trajectories to becoming a DBER professional need to be assessed and understood:

> "Graduate education in DBER is itself ripe for further study and exploration. As DBER fields mature, a growing number of researchers have been trained in DBER graduate programs and are now in academic positions. Now is the time to ask questions, not only about the outcomes of a DBER graduate education (job placement, research productivity/contributions, etc.), but also about best practices for educating graduate students in DBER. These studies would be valuable additions to the literature, and could help to guide the development of programs in newer fields such as astronomy, biology, and geoscience education." (Ref. [2], p. 34).

This article responds to the National Research Council's call by presenting quantitative results from a mixed-methods study focusing on the experiences and educational trajectories of graduate students in physics education research (PER) [1,6].

## II. RESEARCH GOALS: STAGES OF EDUCATIONAL PATH ASSESSED

Our research explores three stages of graduate students' educational trajectories within PER. These stages are (1) Pre-graduate school, (2) graduate school, and (3) future plans. Stage (1) includes discovery of and subsequent choice of pursuing PER, stage (2) includes issues surrounding graduate research topics, and stage (3) includes career goals and expectations of graduate students in PER.

## III. BACKGROUND

Within the literature, few articles currently exist that address the needs and issues of the DBER community. Two such articles in this area of research focus on the lives of faculty [2,3], with a third focusing specifically on geosciences education [4]. Bush et al. surveyed science faculty with education specialties (SFES) about their academic experiences ($N = 289$ respondents) [3]. SFES were defined as faculty in disciplinary departments that focus on education research. Their findings indicated that the phenomena of SFES was on the rise, with more hires occurring in the last decade than all previous decades combined. They also found that the educational training of SFES vary between institution types. For example, faculty at masters level institutions were more likely to have science education training than faculty at PhD granting universities. They also found that faculty at PhD granting institutions were more likely to be nontenure track faculty and more likely to receive education research funding. Additionally, they reported that SFES faculty at PhD granting institutions spend more time on teaching than non-SFES faculty.

Bush et al. argue that the role of science education training in the preparation of SFES faculty is currently unknown since faculty with training were no more likely to receive education research funding than those without. Formal science education training was defined as a post-doctoral position, PhD, or MS. This paper, however, was delimited to not include DBER faculty in nondisciplinary departments. Consequently, their results do not apply to faculty in other departments or schools, such as education. This could be problematic considering many DBER specialists exist outside of science departments. This highlights a particular a need within the literature.

Another study in geoscience education focused on the qualitative experiences of five scholars in geosciences education [4]. The interviews from this study revealed challenges and positive aspects in the geoscience community. In the interviews, the faculty members reported concerns of isolation and constant struggles of not being seen as legitimate researchers in the geoscience community. They did, however, believe this perception was slowly changing for the better.

What is clear from these papers is that the DBER presence is poorly defined in science departments. The roles and duties taken on by DBER faculty may not be well understood, and their academic legitimacy may even be in question. Faculty in at least one discipline face particular challenges and concerns. This concern can be seen as a broader issue when considering results from another paper focusing on the University of California system that indicated 40% of SFES faculty considered leaving their institution [2]. However, without a comparison group it is difficult to say what this statistic means, but it is still interesting to consider because such a high number of DBER faculty considering leaving may be telling of a climate issue of their experience. If the research suggests the role of DBER faculty is poorly defined and they suffer from isolation, their desire to leave may primarily derive from negative departmental experiences. In total, these papers primarily focused on faculty experiences and did not place emphasis on their graduate preparation and experiences.

When specifically considering the field of PER, only one article has been published. The PER article by Barthelemy, Van Dusen, and Henderson [6] was the first of a series of articles on graduate PER student experiences. Our article presented here is a companion piece to the Barthelemy, Van Dusen, and Henderson [6] article that examined the qualitative portions of our mixed methods study. The qualitative portion of the research was an exploratory study using in-depth interviews with 13 graduate students and post-doctoral scholars in PER. The results of the paper focused on the choices of the participants to pursue PER, their experiences within PER, and their future career goals.

In the Barthelemy, Van Dusen, and Henderson [6] study, seven of the thirteen participants chose PER while applying to graduate school while six switched to PER as graduate students or for their post-doctoral positions. As members of the PER community the participants described overall positive interactions with their peers, advisors, and faculty within their departments. This was coupled with six participants reporting experiences of hostility towards the legitimacy of PER as a field of physics. Last, the participants in this study primarily wanted to pursue careers in tenure track positions within universities and colleges. Of these, five specified wanting to work at a research intensive university while four wanted to work at a teaching intensive university. The other four participants wanted careers ranging from consulting to being a full time educational researcher.

For the purposes of the research presented here, there are three interesting findings from the existing literature: (1) there is little prior work investigating researcher trajectories within DBER, (2) research on DBER is biased toward faculty in science departments, and (3) many students in graduate PER are switching in rather than applying specifically to PER graduate programs. The following sections will present the methods, results, and conclusions that followed from a quantitative study formed by the results from our initial qualitative assessment. The primary finding from our qualitative paper that informed this work was the tendency of PER graduate students to switch into the field of PER during graduate school. This finding suggested that, in order to understand the needs of the PER community, closer attention needed to be paid to the trajectories of graduate students in PER.

## IV. METHODS

The data analyzed in this research were collected through an online survey administered to all known graduate students in PER. This section will describe the development and implementation of the survey.

### A. Sampling

To survey the PER graduate student community, we were granted access to the graduate student Email list from the Physics Education Research Consortium of Graduate Students (PERCoGS), a PER graduate student representative body. This list was created to facilitate the first election of PERCoGS and was populated through three iterations: (1) collection of Emails at the American Association of Physics Teachers Summer 2012 Graduate Student Crackerbarrel, (2) Emailing students from the Crackerbarrel asking them to further populate the list which was posted online, and (3) Emailing the Physics Education Research Topical Group to help further complete the list. This final list contained the names and email addresses of 182 graduate students in PER from around the world. The list included any student who identified their graduate research track as being in PER or were identified by an advisor or colleague. While students were primarily based in physics departments and schools of education; student selection was not limited to particular academic programs.

Before the survey was distributed, three senior leaders in the PER community sent a joint introductory Email endorsing the survey and urging graduate students to participate. One day after this Email, the survey was launched. Reminders were sent to nonrespondents every five days until the graduate student either completed the survey or they received three reminders. The survey was open for a total of twenty days and was closed when responses stagnated after the last reminder. In all, there were 125 respondents from the 182 students sampled, giving a response rate of 68%.

### B. Survey instrument

The survey instrument was developed through a three-iteration process: (1) Analysis of graduate student interviews, (2) peer evaluation, and (3) think-aloud interviews. The first iteration was created from the results of interviews conducted in a qualitative study of graduate students and post-doctoral research associates in PER [1]. This draft identified four primary areas for investigation: demographics, trajectories, climate experiences, and motivation of graduate students in PER. The second iteration created by incorporating the collective feedback of the PER research group at the University of Colorado Boulder. In this stage, each survey question was discussed for relevance and purpose to the goals of the study. The third iteration incorporated feedback from a think-aloud process. Two students pursuing graduate research in PER, one in a department of physics and one in a school of education, took the survey and spoke aloud their interpretations and understanding of the questions.

The survey instrument had a total of 51 questions (see Appendix). Not all participants answered all questions, as some had built in logic to only be displayed if relevant to the participant. Included in the survey was a reliability question to test for the participant's close reading and attention to the survey. This question resulted in the removal of one participant's answers. The final survey instrument can be found in the Appendix.

### C. Analysis

The analysis in this paper focuses only on the survey questions surrounding student trajectories. Because of potentially significant difference between countries and the makeup of our data set, the analysis in this paper was restricted to US-based PhD students ($N = 86$). The survey responses were analyzed for connections between how and when they chose to study PER, their current area of research, and their plans for the future. To assess how students ended up choosing to become PER researchers, they were asked four questions: (1) What was your undergraduate degree in? (2) When in your academic career did you discover the field of PER? (3) How in your academic career did you discover PER? (4) When did you choose to study PER at the graduate level? Most of the questions were multiple choice, offering several common answers as well as an "other" option that allowed them to write in an answer.

To understand the PER doctoral students' current environments and future plans, they were asked four additional questions: (1) What department will confer the degree you are seeking? (2) What is the instructional area of your primary research? (3) What best describes where you want to be working in 10 years? (4) What best describes where you think you will be working in 10 years? Again, each of these questions were multiple choice with an "other" option that allowed participants to write in their own answer.

To analyze the data a three-phase process was employed: (1) First, descriptive statistics were used to characterize the population as a whole. Then, in order to understand the relationship between each step within a student's trajectory, (2) we tracked each individual student's path chronologically and created descriptive statistics about these paths. Finally, (3) in order to examine if there were statistically significant differences between these paths, two different types of chi-squared analyses were performed. Because our data are paired and nominal, McNemar's chi-square tests were used when both the dependent and independent variables were binary. When the independent or dependent variables were not binary Pearson's chi-square tests were used. The McNemar's tests were paired with a measure of effect size using the Phi coefficient. The Pearson's chi squares were paired with a measure of effect size using Cramér's V. For all of the analyses related to the flow charts, each row was used to define the groups and the row beneath them defined the potential outcomes. Several steps were shown to be statistically significant, despite our relatively small sample size.

## V. RESULTS

The results will be presented in three stages. First, we will examine students' pasts and how they came to be physics education research graduate students. Second, we will examine the current situations of the PER graduate students. Third, we will examine the plans that the graduate students have for their future career paths.

### A. Stage 1: Pre-graduate school

The paths by which graduate students decided to become physics education researchers are shown in Fig. 1. The majority of PER graduate students (84%) have a degree in physics or astrophysics. Of the 16% of our sample without a physics degree, the students held engineering (3%, $n = 3$), biology (2%, $n = 2$), chemistry (2%, $n = 2$), education (2%, $n = 2$), math (2%, $n = 2$), economics (1%, $n = 1$), physical science (1%, $n = 1$), and psychology (1%, $n = 1$) degrees. The majority of PER graduate students (58%) did not know that PER existed until they were already in graduate school. Those who held an undergraduate physics degree were more likely to have learned of PER's existence as an undergraduate (31% of those with physics degrees and 19% of those without physics degrees). The differences between PER graduate students who did and did not hold undergraduate physics degrees and when they found out about PER (represented in the first and second rows of Fig. 1) are statistically significant ($p < 0.001$) and have a small effect size of 0.140. Of the PER graduate students who learned of PER's existence prior to starting their graduate degree, only 5% of them waited until they were in graduate school to choose to research PER (Table I). Differences in when people who knew of PER prior to graduate school, or not, chose to engage in PER (represented by the second and third rows of Fig. 1) were statistically significant ($p < 0.001$) and have a medium effect size of 0.678. Of the PER graduate students in physics departments, a slight majority of them (53%) decided to study PER after starting their graduate degree. In the schools of education, only 32% of the students decided to study PER after starting their graduate degree. Physics departments have the majority of the PER graduate students (57%), with schools of education making up the other significant contributor (41%). The departmental difference in when PER graduate students chose to engage in PER (represented in the third and fourth rows of Fig. 1) are not statistically significant. To simplify Fig. 1, the 2% (2 students) of students who are not in either a physics department or a school of education are not shown in the final stage.

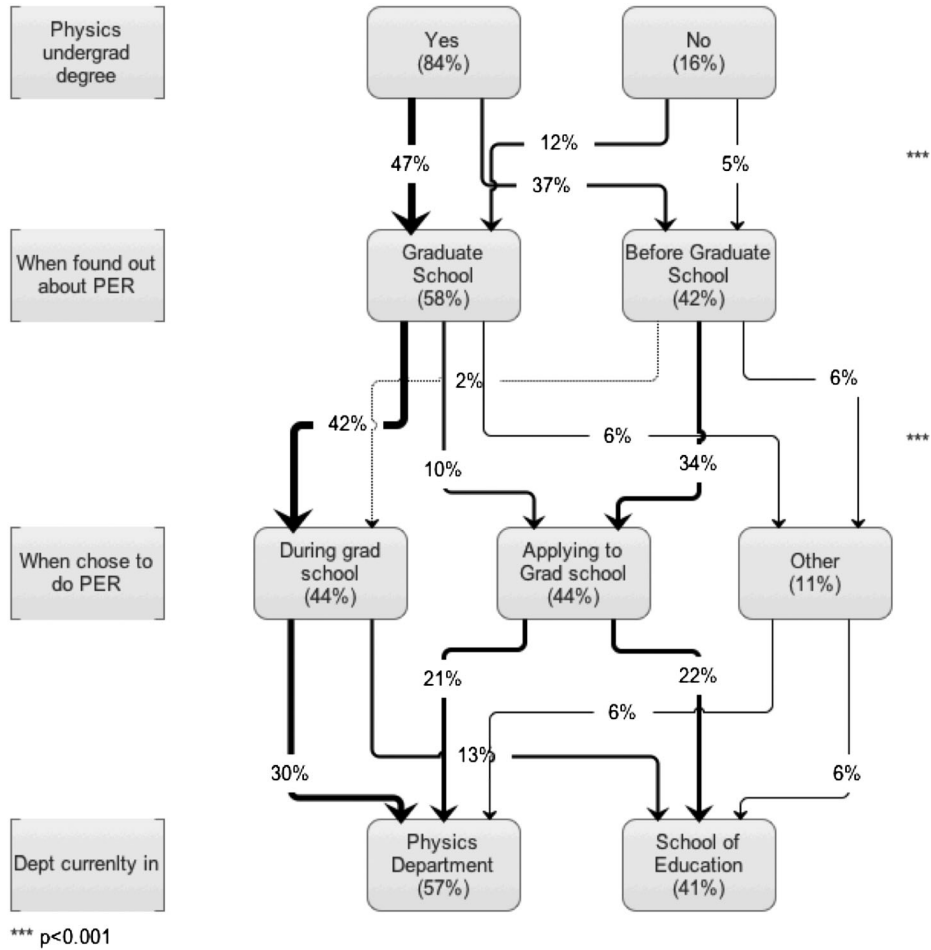

FIG. 1. Paths by which graduate students decided to become PER researchers (values are the percentage of the total population).

TABLE I. When student's found out about PER and when they decided to pursue it (values are the percentage of the subgroup populations).

| When found out about PER | When decided to pursue PER | | |
| --- | --- | --- | --- |
| | Applying to graduate school | During graduate school | Other |
| Graduate school | 17% | 72% | 10% |
| Before graduate school | 81% | 5% | 14% |

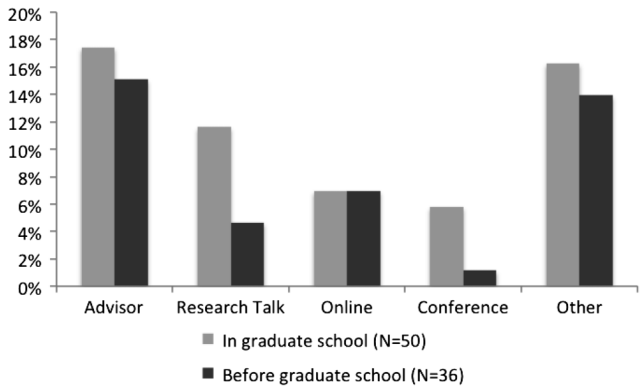

FIG. 2. How students found out about PER (values are the percentage of the total population).

TABLE II. PER graduate students' primary areas of research (values are the percentage of the subgroup populations).

|  | K–12 | Undergrad | Grad., informal, other |
|---|---|---|---|
| Physics ($N = 48$) | 4% ($n = 2$) | 89% | 6% ($n = 3$) |
| Education ($N = 35$) | 37% | 43% | 21% |

Figure 2 shows how students found out about the field of PER. While advisors were the single most common way that students discovered PER, it only represents 32% of the total responses. With no other single source contributing more than 20% of the total, it appears that there is significant diversity in how students came to learn about PER.

### B. Stage 2: Graduate school

The department in which PER graduate students are currently enrolled and their primary areas of research are shown in Fig. 3 and Table II. Examining the primary areas of research shows that the percentage of PER graduate students investigating undergraduate teaching and learning (70%) is more than three times that of the next most popular area of research (K–12). Table II shows that within physics departments, this divide is further accentuated with 89% of the PER graduate students investigating undergraduate settings. Table II also shows that PER graduate students in schools of education have a much more even distribution of primary areas of research (43% undergraduate, 37% K–12, and 21% graduate, informal, or other). Of the 19% of the population primarily investigating K–12 settings, 88% of them are enrolled in schools of education. Departmental differences in the primary area of research for PER graduate students are statistically significant ($p < 0.001$) and have a medium effect size of 0.550.

### C. Stage 3: Post-graduate school

Where PER graduate students *want* to be working in ten years and where they *expect* they will be working in ten years are shown in Fig. 4 and Table III. While the departmental differences between where PER graduate students want to work are not statistically significant, the difference in where they expect to work are statistically significant ($p = 0.032$) and had a small effect size of 0.322. The majority of graduate students (59%) reported wanting the same type of position they expected to be holding in 10 years. Being an academic at a teaching intensive college or university was both the most common place to want to work (34%) and the most common place to expect to work (27%). A similar percentage of PER graduate students in physics departments and schools of education want a teaching intensive academic position (30% in physics departments and 34% in schools of education). Both groups also show a similar percentage of people who expect they will have a teaching intensive academic position (25% in physics departments and 29% in schools of education).

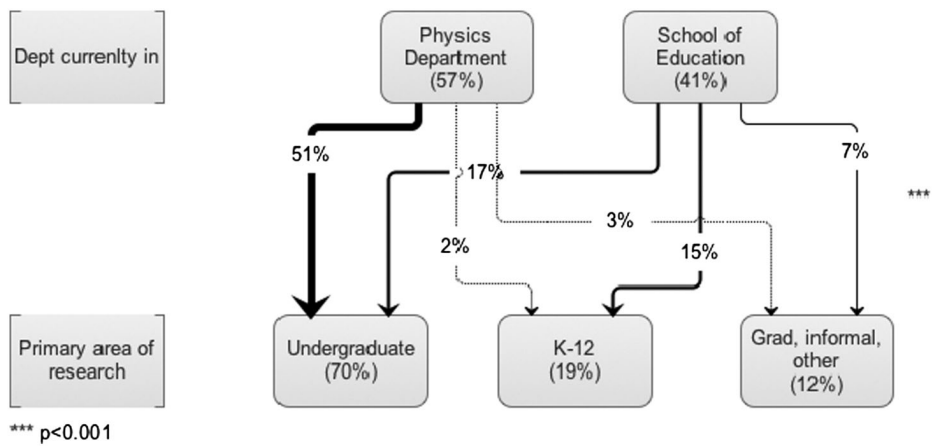

FIG. 3. PER graduate students' primary areas of research (values are the percentage of the total population).

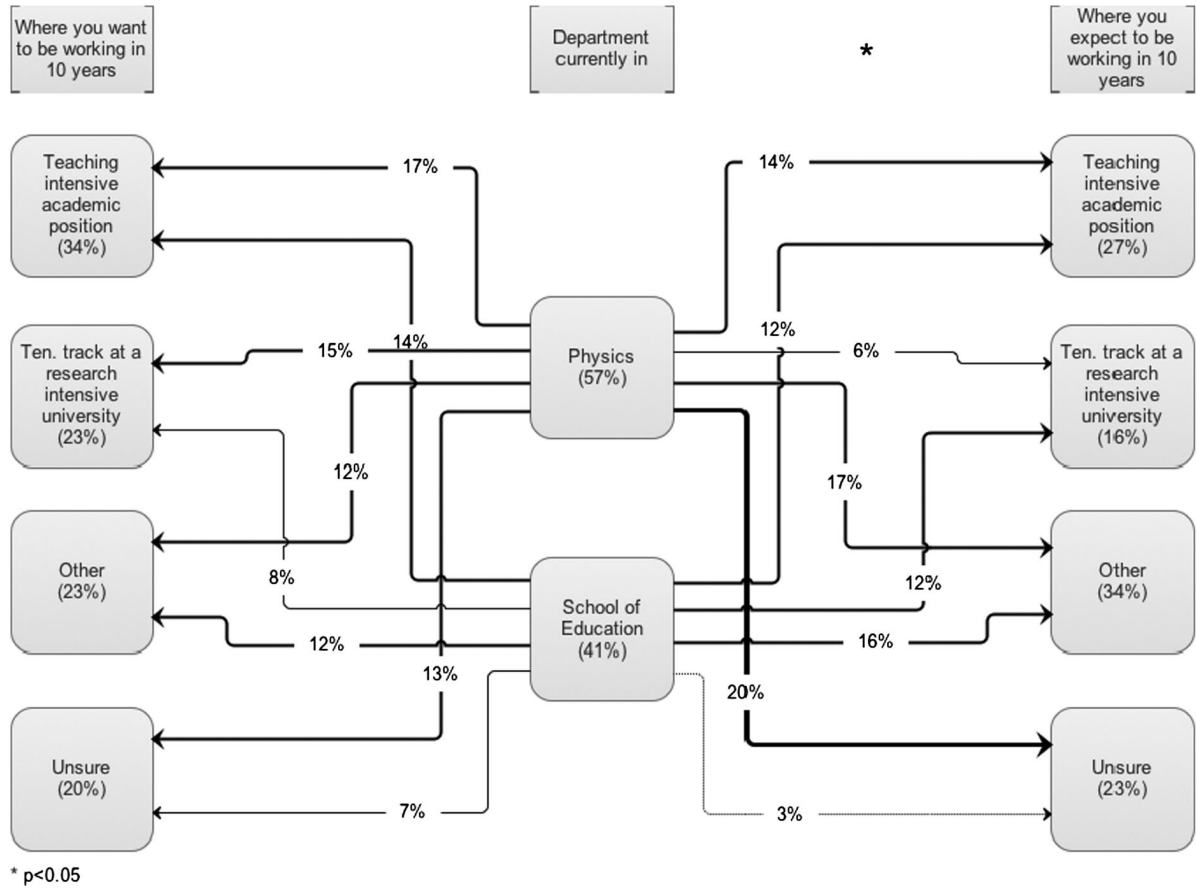

FIG. 4. Where PER graduate students want to work and expect they will be working in ten years (values are the percentage of the total population).

Examining PER graduate student desires and expectations for holding a tenure track position at a research intensive university tells a different story. PER graduate students from physics departments were slightly more likely than ones from schools of education to *want* to hold a tenure track position at a research intensive university (26% of those in physics departments and 20% of those in schools of education). However, the percentage of PER graduate students in physics departments that *expect* they will have a tenure track position at a research intensive university is much lower than PER graduate students in schools of education (11% of those in physics departments ($n = 5$) and 29% of those in schools of education). This shift from desires to expectations can be seen in Fig. 5.

We see a similar, but opposite, shift in the percentage of PER graduate students reporting being unsure about where they want and where they expect to be working in ten years (Fig. 6). Students from physics departments reported slightly higher levels of being unsure about where they *want* to be working than students in schools of education (23% in physics departments and 17% in schools of education). This difference is significantly increased when we compare students reporting being unsure of where they *expect* to work in ten years [35% in physics departments and 7% in schools of education ($n = 2$)].

We examined several potential causes of the differences in PER graduate student levels of uncertainty about their future. The existing career choice literature identifies a variety of factors that play a role in the trajectories of college students [7,8]. Some of the most influential factors to career choices that have been identified include personal interest, work-relevant experiences, and financial barriers. An individual's confidence of their future employment in

TABLE III. Where PER graduate students want to work and expect they will be working in ten years (values are the percentage of the subgroup populations).

|  |  | TT research intensive position | TT teaching intensive position | Other | Unsure |
|---|---|---|---|---|---|
| Physics | Want | 26% | 30% | 21% | 23% |
| ($N = 48$) | Expect | 11% | 25% | 30% | 35% |
| Education | Want | 20% | 34% | 29% | 17% |
| ($N = 35$) | Expect | 29% | 28% | 38% | 7% |

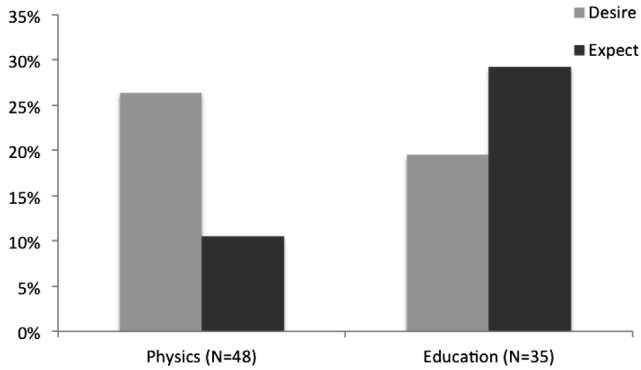

FIG. 5. Percentage of PER graduate students reporting wanting and expecting to hold a tenure track position at a research intensive university (values are the percentage of the subgroup the populations).

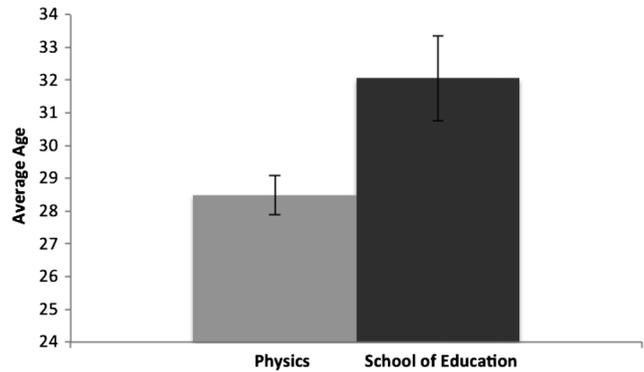

FIG. 7. Average age of PER graduate students in physics departments and schools of education (with error bars representing the standard error of the mean).

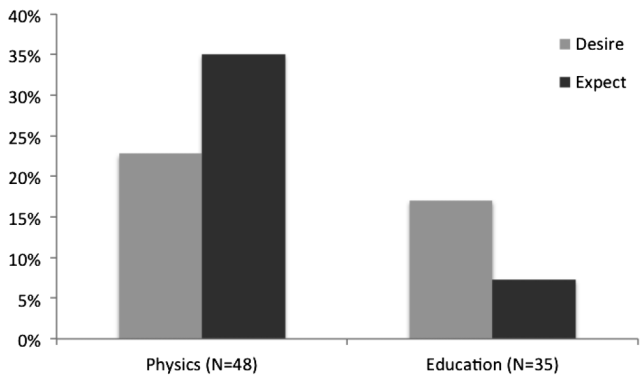

FIG. 6. Percentage of PER graduate students reporting being unsure about where they want to and where they expect to be working in ten years (values are the percentage of the subgroup populations).

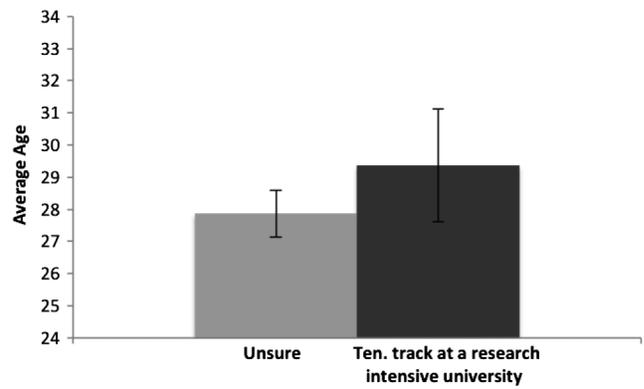

FIG. 8. Average age of PER graduate students who report being unsure of where they expect to be working and those who expect to have a tenure track position in a research intensive university (with error bars representing the standard error of the mean).

an area may be driven in part by their interest in the field, the number and strength of their relevant professional experiences, and lowered barriers to entry (financial and otherwise). While our data do not allow us to directly measure the previously identified variables, we hypothesize that age may be positively correlated with work-relevant experiences and decreased financial barriers. As Fig. 7 shows, PER graduate students in schools of education, who are more confident of their career expectations, are older, on average, than their physics department counterparts

(32.1 years old for schools of education and 28.5 years old for physics departments). However, when examining the average ages of those who report expecting to be working in a tenure track position at a research intensive university with those who report being unsure where they will be working, the difference in age is decreased (29.4 years old for those expecting to be tenure track and 27.9 years old for those who are unsure where they expect to be working) (see Fig. 8). An independent sample $t$ test shows that the difference in the ages of PER graduate students was statistically significantly when separated by departments ($p = 0.005$), but not when separated by job expectation ($p = 0.654$).

TABLE IV. "Other" professional wants and expectations for 10 years in the future (values are the percentage of the total population).

|  |  | Non-TT university position | Informal science education | Science policy | Teaching HS | Education industry |
|---|---|---|---|---|---|---|
| Physics | Want | 2% | 6% | 0% | 4% | 2% |
| ($N = 48$) | Expect | 15% | 2% | 0% | 4% | 2% |
| Education | Want | 14% | 3% | 3% | 0% | 0% |
| ($N = 35$) | Expect | 23% | 3% | 6% | 3% | 0% |

A significant number of PER graduate students expected to be holding positions that fell into our *other* category. The complete list of data from the other category can be seen in Table IV. Within the other category, the most common profession selected for both desires and expectations was the nontenure track university position. PER graduate students both in physics departments and in schools of education reported higher rates of expecting to be working in a nontenure track university position than reported desiring to work there. The nontenure track university positions were particularly undesirable for PER graduate students in physics departments.

## VI. DISCUSSION

PER is a growing subdiscipline of physics. As our results show, PER is unlike other physics subdisciplines in several ways. Not only are PER graduate students split between two departments (physics and schools of education), but they take trajectories that are also unusual in academia. By examining the experiences of our PER graduate students we hope to provide a better understanding of the current state of our field and to draw conclusions about how to increase the growth and productivity of our field.

### A. Knowledge of PER: Stage 1

It is striking that the majority of PER graduate students did not know about the field of PER until they were in graduate school. Even those PER graduate students who have undergraduate degrees in physics primarily discovered PER during their graduate degree. While our data do not directly address the claim, it seems likely that physics undergraduates who do not eventually choose to go into PER have even lower rates of knowing about the existence of PER. This lack of awareness of PER is likely partially caused by the lack of physics departments with PER groups and the fact that PER is not part of a traditional undergraduate curriculum. In this respect, PER has significant hurdles to increasing its exposure that many other physics subfields lack.

Our field primarily relies on a steady supply of midgraduate degree switchers (those who change from a traditional physics field to PER) to fill out the graduate student ranks. This phenomenon of switching is also how many PER faculty came to the field [5,9]. While these switchers have been the foundation for much of the progress PER has accomplished over the past several decades, by tapping into a larger body of physicists and educators earlier in their career we could see significant growth of our field. Ninety-five percent of the current PER graduate students who knew of the field's existence prior to starting their PhD chose to pursue PER prior to beginning their PhD. This indicates that simply informing undergraduates about the existence of PER might produce a significant shift in our ability to recruit new scholars to the field.

### B. Fields of PER research: Stage 2

Our results showed a significant difference between the primary areas of research for PER graduate students in physics departments and those in schools of education. Students based in physics departments were primarily focused on researching undergraduate settings, while those in schools of education were much more evenly spread across all physics learning settings. As a group, 70% of PER graduate students are focused on researching undergraduate education, while only 19% of them are focused on K–12 education. Of the PER graduate students in physics departments, only 4% ($n = 2$) focus on K–12 areas of research. It should be noted that this division might not be reflective of the graduate student's interests. The research settings could simply be a function of the research of the graduate students' advisors and available funding for assistantships.

### C. Future plans: Stage 3

Our survey results showed a reasonable spread of settings in which PER graduate students would like to be working in ten years. Perhaps the most remarkable result was the mismatch between the desires and expectations of holding a tenure track position at a research intensive university. PER graduate students in physics departments showed that more students want a tenure track position at a research intensive university than expected to actually hold one. In addition, the PER graduate students in physics departments reported higher rates of uncertainty in where they thought they would be working and lower rates in where they wanted to be working. The exact opposite trends were observed in PER graduate students in schools of education. The percentage of PER graduate students in schools of education who expected to work at a research intensive university was higher than the percentage that wanted to work there. The percentage of PER graduate students in schools of education expressing uncertainty in where they expected to be working was lower than the percentage that were unsure where they wanted to be working.

As a potential explanation for this split in the uncertainty in one's future plans, we examined the age of the respondents. Older students will have had more work experience and may be in more financially stable situations, which may lead them be more confident about their future careers. PER graduate students in schools of education are, on average, nearly 4 years older (a statistically significant difference). However, the average age of those who expected to be working at a research intensive university was only 1.5 years older than those who were unsure about where they expected to work (a nonstatistically significant difference). Alternatively, physics departments often offer larger salaries and benefits than school of education, which would potentially offset the financial stability that age might offer. This leads us to conclude that while age may be a contributing factor, it does not appear to fully account for these differences.

There are many potential causes of the differences in reported expectations of holding a tenure track position in a research intensive university. For example, it may be that students are aware of respective diminishing percentages of academic positions that are tenure track and the difficulties inherent in obtaining them. In the 2011 9.1% of PhD graduates in the physical sciences and 36.4% of graduates in education found jobs in academia [10]. It should be noted that these data do not distinguish between positions at teaching intensive and research intensive university settings. Additional research needs to be performed to fully explain this finding.

Finally, we also see that students in both physics departments and schools of education expect to be in nontenure track university positions at rates that are higher than their desires to be in them. For the physics department based students, the majority of the students who reported expecting to be in a nontenure track university position originally reported wanting to work in a tenure track position at either a research or teaching intensive university. For students in schools of education, however, the higher rates of expectations over desires for nontenure track university positions primarily came from students who were unsure where they wanted to be working.

## VII. CONCLUSION

Our analysis of graduate students' pre, current, and future PER experiences and plans leads us to draw three primary conclusions. First, the field of PER lacks visibility. Even within the field of physics, PER remains a research specialty that is largely unheard of by physics undergraduates. This lack of visibility makes the act of recruiting future graduate students to the field more difficult. Not only does this mean that PER faculty have to put more energy into recruiting graduate students, but it also means that they will ultimately have a smaller pool of applicants to choose from.

Until the subdiscipline of PER is more visible within the physics and education communities, it will be difficult to attract the kinds of recognition and legitimacy that are critical to creating environments in which a subdiscipline can flourish. Our field's ability to attract funding, create tenure track positions, and to drive change is largely dependent on how we are viewed by non-PER faculty and agencies. While the rise of STEM centers and DBER communities have created more opportunities for PER researchers to publicize their work, it is clear that the field needs to continue to work on increasing its visibility in the physics and education communities.

Our second conclusion is that the field of PER is segmented, particularly in the area of K–12 PER research. As a field, we should consider how we could bring to bear the expertise of all of the interested parties into each area of PER research. While we believe that it is critical that there are collaborations between PER researchers in physics departments and schools of education we do not think that it should stop there. While our data does not directly address it, we believe that PER research could be strengthened through the involvement of faculty and graduate students from non-PER physics departments, schools of education, learning sciences, and discipline based education research groups at large.

Our final conclusion is that the field of PER would benefit from increased graduate student career guidance. This is particularly true for physics departments, where 23% of the graduate students are unsure where they want to work and 35% are unsure where they expect to end up working. Less than half of the PER graduate students in physics departments who desire a tenure track position at a research intensive university expect to actually hold one. While increasing the number of tenured PER positions in universities will likely help with this problem, we should also be actively advising our graduate students as to what career paths exist for them and what they might find most fulfilling.

Our work highlights the need for additional research into undergraduate awareness of PER, employment trajectories of PER graduates, and the areas that PER graduate students and scholars study. By assessing the level of awareness of PER amongst undergraduate students it will help the community to better understand how to disseminate information about our field, so we do not risk losing talented and interested minds. Employment outcomes also emerged as an issue, particularly for PER students in departments of physics. Producing data on where past PER graduates are currently employed would help to build a community-wide understanding of what careers are possible for PER graduates. It would also help equip advisors with information to support their students.

Last, we recommend that future research expand this line of inquiry to include students and scholars from other DBER fields such as chemistry and biology education. Such efforts should seek to include not only persons from science disciplines but also schools of education.